\documentclass[12pt,dvips]{article}

\usepackage{rotating}
\usepackage{axodraw}
\usepackage{epsfig}
\usepackage{psfig}
\usepackage{color}

\newcommand{\s}{\\ \vspace*{-3.5mm} }

\newcommand{\imag}{\Im {\rm m}}
\newcommand{\real}{\Re {\rm e}}

\begin{document}

\mbox{ } \\[-1cm]
\mbox{ }\hfill KEK--TH--819\\%[-1mm] 
\mbox{ }\hfill KIAS--P02023\\%[-1mm] 
\mbox{ }\hfill TUM--HEP--460/02\\%[-1mm] 
% \mbox{ }\hfill hep--ph/0204200\\%[-1mm] 
% \mbox{ }\hfill \today\\%[-1mm]
\bigskip

\thispagestyle{empty}
\setcounter{page}{0}

\begin{center}
{\Large{\bf Supersymmetric Higgs Boson Decays \\[2mm]
            in the MSSM with Explicit CP Violation}} \\[2cm]
            S.Y. Choi$^1$,\, Manuel Drees$^2$,\, 
            Jae Sik Lee$^3$ and J. Song$^4$ 
\end{center}

\vskip 0.5cm

{\small
\begin{enumerate}
\item[{}] { }\ \ \ $^1${\it Department of Physics, Chonbuk National University,
            Chonju 561--756, Korea}\\[-7mm]
\item[{}] { }\ \  \ $^2${\it Physik Dept., TU M\"{u}nchen, James Franck Str., 
            D--85748 Garching, Germany}\\[-7mm]
\item[{}] { }\ \  \ $^3${\it Theory Group, KEK, Tsukuba, Ibaraki 305--0801,
            Japan}\\[-7mm]
\item[{}] { }\ \  \ $^4${\it School of Physics, Korea Institute for Advanced
            Study, Seoul 130--012, Korea}
\end{enumerate}
}
\bigskip
\bigskip
\vskip 3cm

\begin{abstract}
\noindent
Decays into neutralinos and charginos are among the most accessible
supersymmetric decay modes of Higgs particles in most supersymmetric
extensions of the Standard Model. In the presence of explicitly
CP--violating phases in the soft breaking sector of the theory, the
couplings of Higgs bosons to charginos and neutralinos are in general
complex. Based on a specific benchmark scenario of CP violation, we
analyze the phenomenological impact of explicit CP violation in the
Minimal Supersymmetric Standard Model on these Higgs boson decays. The
presence of CP--violating phases could be confirmed either directly
through the measurement of a CP--odd polarization asymmetry of the
produced charginos and neutralinos, or through the dependence of
CP--even quantities (branching ratios and masses) on these phases.
\end{abstract}
%

%\pacs{PACS number(s): 11.30.Er, 12.60.Jv, 13.10.+q}

\newpage

The experimental observation of Higgs particles is crucial for our
understanding of electroweak symmetry breaking. Thus the search for
Higgs bosons is one of the main goals of future colliders such as the
Large Hadron Collider (LHC) and high energy $e^+e^-$ linear colliders
(LC).  Once a Higgs boson is found, it will be of the utmost
importance to perform a detailed investigation of its properties so as
to establish the Higgs mechanism as the basic way to generate the
masses of the known particles. To this end, precise theoretical
predictions for the main decay channels as well as the production
cross sections are essential.\s

In the Minimal Supersymmetric Standard Model (MSSM), CP--violating
phases of some dimensionful parameters (most of which parameterize the
soft breaking of supersymmetry) cause the CP--even and CP--odd neutral
Higgs bosons to mix via loop corrections \cite{AP,CDL}; the most
important contribution usually comes from the top--stop sector. The
loop--induced CP violation in the MSSM Higgs sector can by itself be
large enough to affect the Higgs phenomenology significantly at
present and future colliders~\cite{AP,EXCP_FC,MC1,CL1,
mucol}. Moreover, these CP--phases can also lead to ``direct'' CP
violation in the couplings of Higgs bosons to superparticles
\cite{mucol}. The impact of such potentially large CP--violating
effects on Higgs boson decays has recently been studied in
ref.\cite{CL1}, where the dominant decays of the charged and neutral
Higgs bosons, into standard model (SM) particles and squark pairs,
were investigated in the context of the MSSM with explicit CP
violation. In this note, we extend these analyses by including the
potentially significant decays of Higgs particles into neutralinos and
charginos. These decays have been studied in detail in the
CP--invariant version of the MSSM in refs.\cite{R3a,DJKZ}. We allow
for CP violation both through loop effects in the Higgs sector, using
a form of the Higgs mass matrix that is applicable for all
combinations of stop mass parameters \cite{CDL}, and through phases in
the chargino and neutralino mass matrices. We find that the CP phases
can significantly alter the branching ratios for these decays;
moreover, they can also lead to the appearance of large CP--odd
polarization asymmetries. \s

As well known \cite{R3a,DJKZ}, Higgs boson decays to neutralinos and
charginos, 
\begin{eqnarray}
H^0_k\,\,\rightarrow \tilde{\chi}_i^0\tilde{\chi}_j^0,\ \
                     \tilde{\chi}_i^+\tilde{\chi}_j^- \qquad {\rm and}\qquad
H^\pm \rightarrow \tilde{\chi}_i^0 \tilde{\chi}_j^\pm,
\label{eq:hinodec}
\end{eqnarray}
could play a potentially important role. Here $k=1,2,3$ labels the
three neutral Higgs bosons of the MSSM, while $i,j=1$--$4$ and $1,2$
for neutralinos and charginos, respectively. If $R-$parity is
conserved and $\tilde{\chi}_1^0$ is the lightest supersymmetric stable
particle (LSP), the $\tilde{\chi}_1^0 \tilde{\chi}_1^0$ final states
are invisible. The other $\tilde\chi \tilde\chi$ modes would also be
accompanied by a large amount of missing energy coming from the
$\tilde{\chi}_i^0$ and $\tilde{\chi}_i^\pm$ decay cascades, which lead
to (at least) two LSPs per Higgs boson decay. If kinematically
allowed, the branching ratios for some of the supersymmetric Higgs
boson decay modes (\ref{eq:hinodec}) will be large, unless the ratio
of vacuum expectation values (vevs) $\tan\beta \equiv \langle h_2^0
\rangle / \langle h_1^0 \rangle \gg 1$; here $h_2$ ($h_1$) is the
Higgs doublet coupling to top (bottom) quarks. If $\tan\beta$ is very
large, the $b$ and $\tau$ Yukawa couplings become large, in which case
the modes (\ref{eq:hinodec}) will be subdominant. We will therefore
focus on a scenario with moderate $\tan\beta$.\s

In order to determine the masses of charginos and neutralinos as well
as their couplings to Higgs particles, we have to specify the higgsino
mass parameter $\mu$ and the $U(1)$ and $SU(2)$ gaugino mass
parameters $M_1$ and $M_2$. Following the notation of ref.\cite{CKMZ}, we
write the chargino matrix as
\begin{eqnarray} \label{eq:cmat}
{\cal M}_C=\left(\begin{array}{cc}
                M_2                &  \sqrt{2}m_W c_\beta \\[2mm]
             \sqrt{2}m_W s_\beta  &             \mu   
                  \end{array}\right).
\end{eqnarray}
Diagonalizing this matrix with the help of two unitary matrices $U_R,
U_L$, i.e., ${\cal M}_{C,{\rm diag}} = U_R {\cal M}_C U_L^\dagger$,
generates the light and heavy chargino states $\tilde{\chi}^\pm_i$
($i=1,2$). Similarly, the neutralino mass matrix
\begin{eqnarray} \label{eq:nmat}
\mbox{ }\hskip 2cm {\cal M}_N=\left(\begin{array}{cccc}
  M_1       &      0      &  -m_Z c_\beta s_W  & m_Z s_\beta s_W \\[2mm]
   0        &     M_2     &   m_Z c_\beta c_W  & -m_Z s_\beta c_W\\[2mm]
-m_Z c_\beta s_W & m_Z c_\beta c_W &     0    &     -\mu        \\[2mm]
 m_Z s_\beta s_W &-m_Z s_\beta c_W &  -\mu    &       0
                  \end{array}\right)\
\end{eqnarray}
is diagonalized by the unitary matrix $N$, ${\cal M}_{N,{\rm diag}} =
N^* {\cal M}_N N^\dagger$, leading to four neutralino states
$\tilde{\chi}^0_i$ ($i=1,2,3,4$), ordered with rising mass. In
eqs.(\ref{eq:cmat}) and (\ref{eq:nmat}) we have used $s_\beta \equiv
\sin\beta$, $c_\beta \equiv \cos\beta$, and $s_W,\, c_W$ are the sine
and cosine of the electroweak mixing angle. In CP--noninvariant
theories, all mass parameters can be complex. However, one can always
find a field basis where the $SU(2)$ mass parameter $M_2$ as well as
the vevs are real and positive. The $U(1)$ mass parameter $M_1$ is
then assigned the phase $\Phi_1$, and the higgsino mass parameter
$\mu$ has the phase $\Phi_\mu$. We will adopt this convention in this
paper.\s

The couplings of Higgs bosons to charginos and neutralinos are
determined by the unitary matrices $U_{L,R}$ and $N$ defined above, as
well as by the orthogonal matrix $O$ relating the weak eigenstates $\varphi_k
\equiv \{a, \phi_1,\phi_2\}$ to the three neutral Higgs boson mass
eigenstates $H^0_k$ ($k=1,2,3$), $H = O^T \varphi$ \cite{AP,CDL}; here $\phi_i
= \sqrt{2} \real h_i^0$ and $a = \sqrt{2} \left(s_\beta \imag h_1^0
+ c_\beta \imag h_2^0 \right)$. Specifically, the vertices relevant
for the decays of neutral Higgs bosons are given by:
\begin{eqnarray}
&& \langle\tilde{\chi}^-_{iR}|H^0_k|\tilde{\chi}^-_{jL}\rangle\equiv 
   g X^L_{k;ij}
  = -\frac{g}{\sqrt{2}}\left(U_{Ri1} U^*_{Lj2} G_{2k}
                            +U_{Ri2} U^*_{Lj1} G_{3k}
		       \right),\nonumber\\
&& \langle\tilde{\chi}^0_{iR}|H^0_k|\tilde{\chi}^0_{jL}\rangle\equiv
  \frac{g}{2} Y^L_{k;ij}
  = -\frac{g}{4}\left[ (N^*_{i3}G_{2k}-N^*_{i4}G_{3k})
		     (N^*_{j2}-t_W N^*_{j1}) + (i\leftrightarrow j)\right],
\label{eq:hccL}
\end{eqnarray}
where we have defined the complex coefficients $G_{2k}= O_{2k}-i
s_\beta O_{1k}$ and $G_{3k}= O_{3k}-i c_\beta O_{1k}$. The corresponding
couplings for right--handed charginos and neutralinos are given by
\begin{equation} \label{eq:hccR}
X^R_{k;ij} = X^{L*}_{k;ji}, \ \ \  Y^R_{k;ij}=Y^{L*}_{k;ji}.
\end{equation}
Similarly, the relevant charged-Higgs--neutralino--chargino vertices
are: 
\begin{eqnarray}
&& \langle\tilde{\chi}^0_{iR}|H^+|\tilde{\chi}^-_{jL}\rangle\equiv
   g Z^L_{ij}
  = -\frac{g\, s_\beta}{\sqrt{2}}\left[\sqrt{2} N^*_{i3} U^*_{Lj1} 
           -(N^*_{i2}+ t_W N^*_{i1}) U^*_{Lj2}\right],\nonumber\\
&& \langle\tilde{\chi}^0_{iL}|H^+|\tilde{\chi}^-_{jR}\rangle\equiv
  g Z^R_{ij}
  = -\frac{g\, c_\beta}{\sqrt{2}}\left[\sqrt{2} N_{i4} U^*_{Rj1} 
           +(N_{i2}+ t_W N_{i1}) U^*_{Rj2}\right].
\label{eq:hcn}
\end{eqnarray}
The couplings of eqs.(\ref{eq:hccL}) and (\ref{eq:hcn}) show that all
the Higgs particles couple to one gaugino and one higgsino component
of charginos and neutralinos. This is not surprising, since these
interactions result from the supersymmetric Higgs
boson--gaugino--higgsino interactions in the basic supersymmetric
Lagrangian written in terms of current eigenstates. In particular, in
the limit of vanishing higgsino--gaugino mixing, i.e., $|\mu|
\rightarrow \infty $ or $|M_{1,2}| \rightarrow \infty$, all diagonal
$H \tilde\chi_i \tilde\chi_i$ couplings vanish at the
tree--level.\footnote{Under certain circumstances, non--negligible
diagonal couplings can result even in the absence of gaugino--higgsino
mixing once loop corrections have been included \cite{couploop}.} On
the other hand, when $|\mu| \sim |M_1|$ or $|\mu| \sim |M_2|$,
gaugino--higgsino mixing will be sizable and the $H \tilde\chi_i
\tilde\chi_i$ couplings can be significant. Moreover, the total decay
width for Higgs boson decays into charginos and neutralinos will
remain large even for small higgsino--gaugino mixing, if the Higgs
mass in question exceeds $|M_2| + |\mu|$ and $|M_1| + |\mu|$, so that
decays into one gaugino--like and one higgsino--like state are
allowed. 

CP violation in the couplings of neutral Higgs bosons to CP
self--conjugate final states is signaled by the simultaneous existence
of scalar and pseudoscalar components, which happens if, e.g., neither
$X^R = X^L$ nor $X^R = -X^L$. Eq.(\ref{eq:hccR}) therefore implies
that CP will be violated unless the couplings $X,Y$ are either purely
real or purely imaginary. From eq.(\ref{eq:hccL}) we see that such a
nontrivial phase in the couplings results if either the mixing
matrices $U_L, U_R, N$ are complex, due to CP violation in the
chargino and neutralino sector, or if $O_{1k}$ and $O_{(2,3)k}$ are
simultaneously nonzero for some Higgs boson $H_k^0$. The latter signals
``indirect'' CP violation through mixing between scalar and
pseudoscalar Higgs fields.  In the MSSM this mixing is predominantly
induced by loops involving top squarks, and is quantified by the
dimensionless parameter \cite{CDL}
\begin{eqnarray} \label{eq:mixpar}
\Delta_{\tilde{t}}
  =\frac{\imag(A_t \mu){ }}{m^2_{\tilde{t}_2}-m^2_{\tilde{t}_1}}.
\end{eqnarray}
This mixing will be large only if $\imag(A_t \mu)$ is comparable to
the squared top--squark masses.\footnote{Eq.(\ref{eq:mixpar}) seems to
imply that $\imag(A_t \mu)$ only has to be comparable to the {\em
differences} of the squared stop masses. However, the mixing between
(nearly) degenerate states, while apparently large, has no physical
effect.} Finally, the contributions from the top (s)quark sector to
the CP--even Higgs boson masses depend on the magnitude of the top
squark mixing parameter $X_t= - m_t(A_t+\mu^*/ \tan\beta)$ as well as
on the soft--breaking top--squark mass parameters, $m_{\tilde{Q}}$ and
$m_{\tilde{t}}$. So, at the one--loop level the Higgs boson masses (in
particular, $m_{H_1}$) will depend significantly on the rephasing
invariant phase $\Phi\equiv {\sf arg}(A_t\mu)$ only when $|A_t|$ and
$|\mu|/\tan\beta$ are comparable in size. However, our treatment
\cite{CDL} includes leading two--loop corrections by using
appropriately one--loop corrected top quark masses in the loop
corrections to the Higgs boson masses. The gluino--stop loop
corrections to $m_t$ introduce some dependence of the Higgs boson
masses on CP--violating phases even if $|A_t| \gg |\mu|$. Our
calculation thus includes pure Yukawa and mixed electroweak
gauge--Yukawa corrections to one loop order exactly (using the
effective potential method), as well as leading (SUSY)QCD two--loop
corrections. However, we do not include purely electroweak loop
corrections to the Higgs masses \cite{IN}.\s

Motivated by the above observations and experimental constraints on
the lightest Higgs boson mass and on the light chargino mass
\cite{PDG}, we consider the following benchmark scenario of SUSY
parameters\footnote{A phase of the $SU(3)$ gaugino mass parameter
$M_3$ could modify the top-- and bottom--quark Yukawa couplings at the
one loop level, and could thus affect the branching ratios of the
supersymmetric decays \cite{Hbb}.}:
\begin{eqnarray}
&& \tan\beta=5;\quad M_A = 0.3\,\, {\rm TeV}; \quad
m_{\tilde{Q}}=m_{\tilde{t}}=0.5\,\, {\rm TeV},  
\quad |A_t|=1.2\,\, {\rm TeV},\quad |\mu|=250\, {\rm GeV}, \nonumber\\
&& |M_1|=50\,\, {\rm GeV}, \quad M_2=150\,\, {\rm GeV};\quad
   |M_3|=0.5\, {\rm TeV}, \quad \arg(M_3)=0.
\label{eq:MCPX}
\end{eqnarray}
Here $M_A$ is the RG--invariant one--loop improved pseudoscalar mass
parameter, which sets the scale for the masses of the heavy MSSM Higgs
bosons. Our choice $M_A = 0.3$ TeV implies that all these Higgs bosons
will be accessible to the second stage of future LCs as currently
planned, which will reach cms energy $\sqrt{s} \sim 0.8$ to $1.2$
TeV. Since in many SUSY models squark and Higgs boson masses are
correlated, we also took relatively modest values for the soft
breaking masses $m_{\tilde Q}, m_{\tilde t}$ of $SU(2)$ doublet and
singlet stops, respectively. One then needs $|A_t| \sim \sqrt{6}
m_{\tilde Q}$ (the so--called maximal stop mixing scenario) in order
to safely satisfy the experimental lower bound on $M_{H_1}$; note that
$H_1^0$ behaves similar to the SM Higgs boson in our case. A large
$|A_t|$ also tends to maximize the CP--violating mixing in the Higgs
sector. Our choice of gaugino and higgsino mass parameters ensures
significant mixing between $SU(2)$ gauginos and
higgsinos. Furthermore, the gaugino masses are sufficiently small that
the first two neutralinos and the lighter charginos are accessible to
the decays of the heavy Higgs bosons, while $H_1^0$ can at least decay
into two LSPs; note that the branching ratio for this last decay can
be sizable only if the gaugino mass ``unification condition'' $M_1
\simeq M_2/2$ is violated \cite{french}. On the other hand, Higgs
boson decays into one gaugino--like and one higgsino--like state,
which have the potentially largest branching ratios of all decays
(\ref{eq:hinodec}), are {\em not} allowed kinematically in our
scenario. Moreover, our value of $\tan\beta$ is neither very large nor
very small. We therefore consider our choice of parameters to be quite
representative of general MSSM scenarios.\s

We have not yet fixed the values of most CP--violating phases. There
are important constraints on these phases in the MSSM, from
experimental limits on the electric dipole moments (EDM) of the
electron, neutron and $^{199}{\rm Hg}$ \cite{EDM}. However, these
constraints can be avoided if there are cancellations between
different supersymmetric diagrams and/or between different
CP--violating operators. Furthermore, since the constraints apply
essentially only to the first and possibly second generation of matter
fermions, they may be more relaxed for the third--generation coupling
$A_t$, if we do not impose the assumption of universality between
different generations. Large phases of $\mu$ and $M_1$ are also
allowed if first and second generation sfermions are much heavier than
sfermions of the third generation. We will therefore consider the
entire range of $\Phi, \Phi_\mu$ and $\Phi_1$ between 0 and $\pi$. \s

\begin{figure}[h]
\vspace*{-1.6cm}
%\hspace*{0.5cm}
\centerline{\psfig{figure=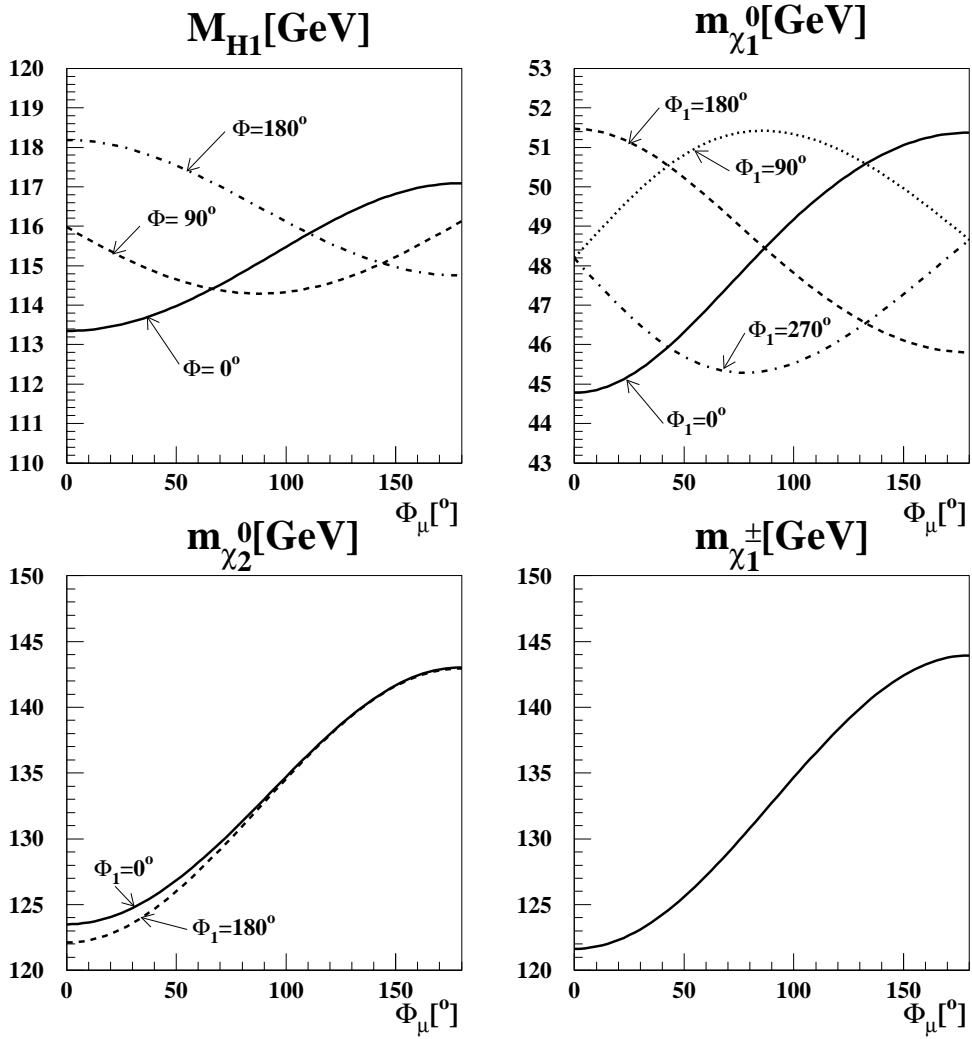,width=15cm}}
\vspace*{-0.5cm}
\caption{\it The lightest Higgs boson mass, the 
         lightest neutralino mass, the second lightest neutralino mass and 
	 the lighter chargino mass as functions of the phase $\Phi_\mu$
	 for various values of $\Phi \equiv \arg (A_t \mu)$ and $\Phi_1$.}
\label{fig:mass}
\end{figure}

Fig.~\ref{fig:mass} shows the dependence of the lightest Higgs boson
mass, the two light neutralino masses and the light chargino mass on
the phase $\Phi_\mu$ for various values of the CP--violating phases
$\Phi$ and $\Phi_1$. Once gluino--stop loop corrections to $m_t$ are
included, the neutral Higgs boson masses depend on both $\Phi$ and
$\Phi_\mu$; indeed, since $|\mu| \cot \beta \ll |A_t|$ in our
scenario, the phase dependence of $m_{H_1}$ comes almost entirely from
these corrections. Since they contribute to the Higgs masses only at
two loop, the maximal variation of $M_{H_1}$ with respect to both
$\Phi$ and $\Phi_\mu$ is less than 5 GeV, $\sim5\%$ of the Higgs mass
itself. We find that the heavy Higgs\\

\noindent boson masses (not shown) also remain almost constant with
their values close to 300 GeV. The lightest neutralino mass shows a
somewhat stronger dependence on both $\Phi_1$ and $\Phi_\mu$ as shown
in the upper right frame of fig.~\ref{fig:mass}. We therefore expect
the branching ratio of the invisible decay $H_1^0\rightarrow
\tilde{\chi}^0_1 \tilde{\chi}^0_1$ to be more sensitive to $\Phi_\mu$
and $\Phi_1$ than to $\Phi$. Moreover, in spite of the large value of
$|A_t|$ we found that CP--violating mixing between the heavy neutral
Higgs bosons amounts to at most a few percent; this is due to the
relatively small value of $|\mu|$. We therefore simply take $\Phi=0$
in the following. The lower frames in fig.~\ref{fig:mass} show that
the approximate equality of $m_{\tilde \chi_2^0}$ and $m_{\tilde
\chi_1^\pm}$ is maintained even in the presence of CP violation.
$m_{\tilde \chi_1^\pm}$ is manifestly independent of $\Phi_1$, but the
$\Phi_1$ dependence of $m_{\tilde \chi_2^0}$ is also essentially
negligible. However, both masses depend quite strongly on $\Phi_\mu$.
For our choice of parameters, both $\tilde \chi_2^0$ and $\tilde
\chi_1^\pm$ are dominantly $SU(2)$ gauginos, with significant higgsino
admixtures.\s

Since charginos and neutralinos are spin--1/2 particles, spin
correlations of the $\tilde\chi \tilde\chi$ pair in the decay $H
\rightarrow \tilde\chi \tilde\chi$ may allow us to probe CP violation
in supersymmetric Higgs boson decays directly. The $\tilde \chi$
momenta cannot be identified event by event, due to the presence of
invisible lightest neutralinos in the final state. Nevertheless,
correlations may be estimated by statistically studying visible decay
products from the spin--correlated $\tilde\chi \tilde\chi$
pairs. Including spin correlations in the final state, we can write
the general form of the spin--correlated widths of supersymmetric
Higgs boson decays into neutralino and chargino pairs in the following
compact form:
\begin{eqnarray}
\Gamma (\vec{P}^i,\vec{P}^j) 
     \,=\, \frac{g^2M_{H_k}\lambda^{1/2}}{16\pi S_{ij}}
     \left\{ C^{ij}_0 (1\!+\!P^i_L P^j_L)
         +\! C^{ij}_1 (P^i_L\!+\!P^j_L)
         + P^i_TP^j_T \left[\,C^{ij}_2\! \cos\phi_{ij}\! 
                            +\!C^{ij}_3\! \sin\phi_{ij}\right]\right\}.
\label{eq:hdec}
\end{eqnarray}
Here $P^{i,j}_L$ and $P^{i,j}_T$ are the degrees of longitudinal and
transverse polarization of the final charginos or neutralinos,
$\tilde\chi_i$ and $\tilde\chi_j$, respectively; $S_{ij}=1$ unless the
final state consists of two identical (Majorana) neutralinos in which
case $S_{ii}=2$; and $\lambda=(1- \mu^2_{ik} - \mu^2_{jk})^2- 4
\mu^2_{ik}\mu^2_{jk}$ with $\mu^2_{ik}=m^2_{\tilde{\chi}_i}/m^2_{H_k}$
is the usual two--body phase space function. Fig.~2 shows a
schematic description of the polarization configuration. The
coefficients $C_i$ ($i=0,1,2,3$) in eq.(\ref{eq:hdec}) are given by
\begin{eqnarray}
C^{ij}_0 &=& (1-\mu^2_{ik}-\mu^2_{jk}) \left( |Q^L_{k;ij}|^2 +
|Q^R_{k;ij}|^2 \right)  
- 4 \mu_{ik} \mu_{jk} \real(Q^L_{k;ij}Q^{R*}_{k;ij}), \nonumber\\
C^{ij}_1 &=& \lambda^{1/2} \left( |Q^L_{k;ij}|^2 - |Q^R_{k;ij}|^2
\right) ,\nonumber\\
C^{ij}_2 &=& 2 (1-\mu^2_{ik}-\mu^2_{jk}) \real \left( Q^L_{k;ij}
Q^{R*}_{k;ij} \right) - 2 \mu_{ik} \mu_{jk} \left( |Q^L_{k;ij}|^2 +
|Q^R_{k;ij}|^2 \right), \nonumber\\
C^{ij}_3 &=& -2 \lambda^{1/2} \imag \left( Q^L_{k;ij} Q^{R*}_{k;ij}
\right).
\label{eq:coefficients}
\end{eqnarray}
Here the general couplings $Q$ stand for $X,Y$ or $Z$ from
eqs.(\ref{eq:hccL})--(\ref{eq:hcn}) for chargino--chargino,
neutralino--neutralino or chargino--neutralino pairs, respectively.\s
\mbox{ }\\[-2.4cm]

\begin{center}
\begin{picture}(500,200)(10,20)
\SetWidth{0.8}

\GCirc(250,100){5}{0}
\Text(252,80)[c]{\Large\color{black} $H_k^0$}
\LongArrow(255,100)(400,100)
\LongArrow(245,100)(100,100)
\Text(95, 100)[r]{\Large\color{black} $\tilde\chi_{_j}$}
\Text(410,100)[l]{\Large\color{black} $\tilde\chi_{_i}$}
\DashLine(190,100)(240,150){9}

\Text(355,135)[c]{\color{blue} $P^i_T$}
\Text(355,86)[c]{\color{blue} $P^i_L$}
\Text(167,140)[c]{\color{blue} $P^j_T$}
\Text(145,86)[c]{\color{blue} $P^j_L$}
\Text(205,140)[c]{\color{red} $\phi_{ij}$}

\LongArrowArcn(190,100)(30,100,45)
\SetWidth{0.5}
\Line(20,50)(110,150)
\Line(390,50)(480,150)
\Line(20,50)(390,50)
\Line(110,150)(480,150)

\SetWidth{2.0}
\LongArrow(310,100)(340,130)
\LongArrow(310,100)(350,100)
\LongArrow(190,100)(180,140)
\LongArrow(190,100)(150,100)
\label{fig:configuration}
\end{picture}
\end{center}
\vskip 0.0cm
{\bf Figure~2}: {\it Schematic description of the longitudinal 
and transverse polarization vectors $P_L^{i,j}$ and $P^{i,j}_T$,
respectively, of the states $\tilde\chi_i$ and $\tilde\chi_j$.
Here, $\phi_{ij}$ is the relative azimuthal angle of $P^i_T$ 
with respect to $P^j_T$.}
\bigskip

Note that all terms in eq.(\ref{eq:hdec}) that are proportional to $
P_L^{i,j}$ or $P_T^{i,j}$ will vanish after summation over $\tilde
\chi$ spins. The various branching ratios are therefore determined
entirely by the $C_0^{ij}$. Eqs.(\ref{eq:hccL})--(\ref{eq:hcn}) show
that the couplings $Q$ all result from adding two or more terms. This
means that not only $\real(Q^L Q^{*R})$ but also the absolute values
$|Q^L|, \ |Q^R|$ are sensitive to the CP--violating phases. The partial
widths also depend on the CP--violating phases through the masses of
the Higgs bosons, charginos and neutralinos.\s

\vskip -2.9cm
\addtocounter{figure}{1}
\begin{figure}[h]
\hspace*{0.5cm}
\centerline{\psfig{figure=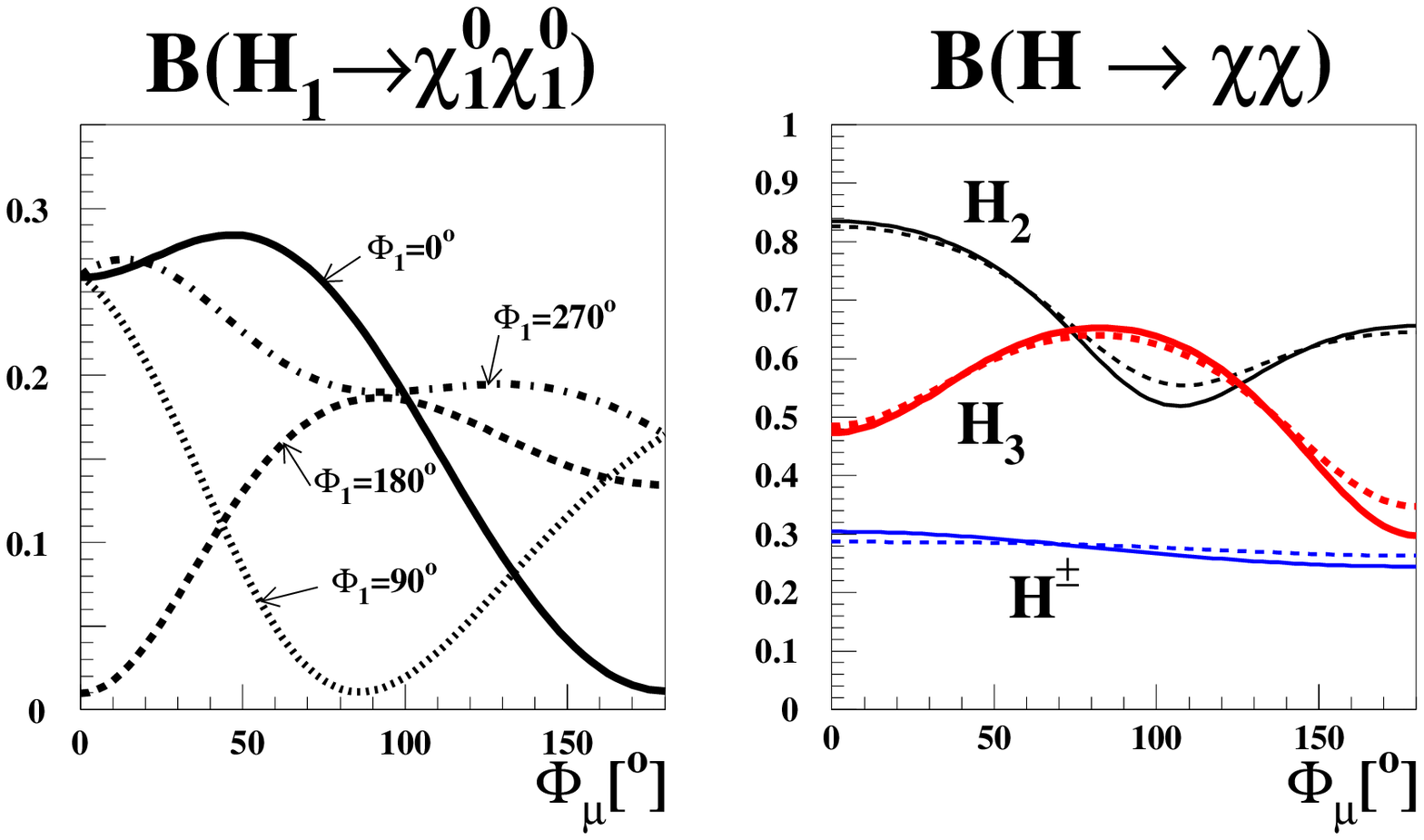,width=15cm}}
\vspace*{-3.3cm}
\caption{\it The branching ratio of the lightest Higgs boson decay into 
         the lightest neutralino pair as a function of $\Phi_\mu$ for 
	 $\Phi = 0$ (i.e., $\Phi_{A_t} =-\Phi_\mu$), and
         $\Phi_1=0^\circ, 90^\circ, 180^\circ$ and $270^\circ$, 
         respectively (left frame) and 
	 the sum of the branching ratios of the heavy Higgs bosons, 
	 $H_{2,3}^0$ and $H^\pm$ decays into all possible chargino and 
	 neutralino modes as a function of $\Phi_\mu$ (right frame). 
	 In the right frame the solid lines are for $\Phi_1=0$ and 
	 the dotted lines for $\Phi_1=180^\circ$.}
\label{fig:brh}
\end{figure}

This is illustrated in fig.~\ref{fig:brh}. The left frame shows the
branching ratio of the invisible decay $H_1^0\rightarrow\tilde{\chi}^0_1
\tilde{\chi}^0_1$. Such decays can be detected quite straightforwardly
at $e^+e^-$ colliders by measuring the missing mass in $Z H_1^0$
events \cite{teslatdr}. It has recently been argued that a measurement
of this invisible branching ratio of $H_1^0$ with an accuracy of a few
percent should also be possible at the LHC, using $H_1^0$ produced in
$WW$ and $ZZ$ fusion \cite{virus}. We see that this decay rate is very
sensitive to $\Phi_1$ and $\Phi_\mu$. This is partly due to the
dependence of the lightest neutralino mass on the phases, see
fig.~\ref{fig:mass}; note that the phase space for this decay is quite
small, so that relatively minor variations of the mass translate into
large changes of the branching ratio. Moreover, this decay is $P-$wave
suppressed, i.e., the partial width is $\propto \lambda^{3/2}$, if CP
is conserved, but develops an $S-$wave piece in the presence of
CP violation; for example, $C_0^{11}$ in eq.(\ref{eq:coefficients})
vanishes at threshold ($\mu_{1k} = 0.5$) in the absence of
CP violation. The branching ratio is therefore maximal at
non--trivial values of $\Phi_\mu$ and/or $\Phi_1$. It is suppressed
near $\Phi_\mu+\Phi_1=180^\circ$ [mod $360^\circ$], where
$m_{\tilde{\chi}^0_1}$ is maximal as shown in the upper right frame of
fig.~\ref{fig:mass}.\s

The sum of the branching ratios for the heavy neutral and charged
Higgs boson decays into all possible neutralino and chargino modes is
shown in the right frame of fig.~\ref{fig:brh}. In our case the four
decay channels $\tilde{\chi}^0_1\tilde{\chi}^0_1, \tilde{\chi}^0_1
\tilde{\chi}^0_2, \tilde{\chi}^0_2\tilde{\chi}^0_2$ and
$\tilde{\chi}^+_1 \tilde{\chi}^-_1$ are allowed for the neutral Higgs
bosons $H_{2,3}^0$, while the two channels $\tilde{\chi}^-_1
\tilde{\chi}^0_1$ and $\tilde{\chi}^-_1 \tilde{\chi}^0_2$ are allowed
for the charged Higgs boson $H^-$. At the LHC the dominant production
process for the heavy neutral Higgs bosons is single production from
gluon fusion, including production in association with a $b \bar b$
pair. It has been shown \cite{atlastdr} that under favorable
circumstances $H^0_{2,3} \rightarrow \tilde \chi_2^0 \tilde \chi_2^0$
decays can be detected at the LHC in the four lepton final
state. However, this requires a large leptonic branching ratio for
$\tilde \chi_2^0$, which in turn requires relatively light
sleptons. We saw above that scenarios with light sleptons and CP
violation are constrained severely by the electric dipole moment of
the electron. Moreover, since the Higgs production cross sections at
hadron colliders are uncertain even in the framework of the MSSM, it
is not easy to translate a measurement of a number of events into a
measurement of the corresponding branching ratio.\s

The dominant heavy Higgs production mechanisms at future $e^+e^-$
colliders \cite{teslatdr} are $H^+ H^-$ and $H_2^0 H_3^0$
production.\footnote{ Note that diagonal $H_i^0 H_i^0$ production
remains forbidden at $e^+e^-$ colliders even in the presence of
CP violation, due to the Bose symmetry of the final state.} The best
search strategy is then probably to look for the decay of one of the
heavy Higgs particles into third generation fermions, while the other
one is required to decay into $\tilde \chi$ states. We are not aware
of a dedicated analysis of such final states, but the presence of an
invariant mass peak for the third generation fermion pair should allow
to extract this signal relatively cleanly. Alternatively one might
simply measure the number of $b \bar b b \bar b$ and $b \bar b \tau^+
\tau^-$ events with double invariant mass peak. Together with
theoretical predictions for the total $H_2^0 H_3^0$ production cross
section, which in the MSSM essentially only depends on $M_A$ once
$M_A^2 \gg m_Z^2$, this would allow to determine the heavy Higgs
bosons' branching ratios into non--SM particles. This could be equated
with the branching ratios for $H_i^0 \rightarrow \tilde \chi \tilde
\chi$ decays if direct searches at the same experiment do not find
other light sparticles into which the Higgs bosons might decay. We
therefore expect the branching ratios to be measurable at future
$e^+e^-$ colliders with rather high accuracy; this should be true at
least for the average of the $H_2^0$ and $H_3^0$ branching ratios,
since it might be difficult to distinguish between these two Higgs
bosons on an event--by--event basis. Finally, the heavy neutral Higgs
bosons can also be produced singly as $s-$channel resonances at future
$\mu^+ \mu^-$ colliders \cite{MC2}.\s

We see that the summed branching ratios of the neutral heavy Higgs
bosons are always quite large, varying between 30\% and 80\% depending
on the value of $\Phi_\mu$. Since we have set $\Phi = 0$, $H_2^0$ is a
pure CP--odd state (often called $A$), while $H_3^0$ is purely
CP--even. Fig.\ref{fig:mass} shows that the phase space for the decays
in question decreases monotonically as $\Phi_\mu$ increases from 0
to $180^\circ$. Nevertheless the $H_2^0 \rightarrow \tilde\chi \tilde
\chi$ branching ratio reaches a minimum for an intermediate value for
$\Phi_\mu$. The reason is that the decay is now purely $S-$wave in the
absence of CP violation, whereas nontrivial CP--phases introduce a
sizable $P-$wave component, which is strongly phase space suppressed
in our case. For example, the $H_2^0 \tilde \chi_1^+ \tilde \chi_1^-$
coupling is almost purely scalar, rather than pseudoscalar, for
$\Phi_\mu \simeq 100^\circ$, near the minimum of $B(H_2^0 \rightarrow
\tilde \chi \tilde \chi)$. The branching ratio of $H_3^0$ decays shows
essentially the opposite behavior, since $H_3^0$ is a CP--even state;
it can decay into an $S-$wave final state only in the presence of
CP violation.

The $\tilde \chi_1^0 \tilde \chi_1^0$ final state is subdominant in
neutral Higgs boson decays; the larger phase space available for it is
over--compensated by the small couplings to this Bino--like
neutralino. The couplings of $H_2^0$ to $\tilde \chi_1^+ \tilde
\chi_1^-$, $\tilde \chi_1^0 \tilde \chi_2^0$ and $\tilde \chi_2^0
\tilde \chi_2^0$ behave similarly, decreasing in magnitude with
increasing $\Phi_\mu$; however, the corresponding couplings of the
CP--even state $H_3^0$, while again similar to each other, show the
opposite dependence on $\Phi_\mu$.  This can be traced back to the
different decomposition of these two heavy Higgs bosons in terms of
current eigenstates: $H_2^0 = \sqrt{2} \imag( s_\beta h_1^0 + c_\beta
h_2^0)$, while for $M_A^2 \gg M_Z^2$, $H_3^0$ is approximately given
by $\sqrt{2} \real ( s_\beta h_1^0 - c_\beta h_2^0)$. In contrast,
$H^-$ decays into neutralinos and charginos are dominated by the
$\tilde \chi_1^- \tilde \chi_1^0$ final state, since the couplings
$Z^{L,R}_{21}$ of eqs.(\ref{eq:hcn}) are suppressed by large
cancellations between the two terms in the square brackets. Note that
the ratio of left-- and right--handed $H^- \tilde \chi_1^- \tilde
\chi_1^0$ couplings is proportional to $\tan\beta$. The charged Higgs boson
decays therefore always have a large $S-$wave component, and are thus
less sensitive to $\tilde \chi$ masses than neutral Higgs decays are;
the $\tilde \chi$ mass dependence is reduced even further since the
phase space for the $\tilde \chi_1^- \tilde \chi_1^0$ mode is anyway
quite large. Furthermore, we find that the absolute value of the
dominant coupling $Z^L_{11}$ depends very little on the phases
$\Phi_\mu$ and $\Phi_1$. The phase of this coupling does vary greatly,
but this has little effect on the absolute value of the coefficient
$C_0$ of eq.(\ref{eq:coefficients}), which determines the
corresponding partial width, since $|Z^R_{11}| \ll |Z^L_{11}|$. Note
finally that the branching ratio for the $\tilde \chi_1^- \tilde
\chi_1^0$ mode is significant even though $H^- \rightarrow b \bar t$
decays are allowed. This indicates that the branching ratios for
$H^0_{2,3} \rightarrow \tilde \chi \tilde \chi$ can also be sizable
even if $M_A > 2 m_t$; recall that the partial widths for $H^0_{2,3}
\rightarrow \tilde \chi \tilde \chi$ decays will increase
significantly if $M_A > |M_2| + |\mu|$.\s

In principle the spin--dependent terms in eq.(\ref{eq:hdec}) allow
more direct probes of CP violation. In case of neutral Higgs boson
decays the $C_0$ and $C_2$ terms are even under a CP transformation
while the $C_1$ and $C_3$ terms are odd. Moreover, the $C_0$, $C_2$
and $C_3$ terms are even under a ${\rm CP}\widetilde{\rm T}$ transformation,
while the $C_1$ term is odd; here $\widetilde{\rm T}$ describes ``naive''
time reversal, which flips the sign of all 3--momenta and spins but
does not exchange the initial and final state. Note that a term can
only be ${\rm CP}\widetilde{\rm T}-$odd but ${\rm CPT}-$even if it depends 
on some
CP--invariant, absorptive phase. No such phase exists in our case (at
the tree--level), so we expect the $C_1$ terms to vanish for neutral
Higgs decays; eqs.(\ref{eq:hccR}) show explicitly that this is indeed
the case. The situation is a bit more complicated for the decays of
charged Higgs bosons, since here the initial and final states are not
CP self--conjugate. However, since spin correlations can only be
measured if both $\tilde \chi$ states produce visible decay products,
while $H^- \rightarrow \tilde \chi^- \tilde \chi^0$ decays are
dominated by the $\tilde \chi_1^- \tilde \chi_1^0$ final state, we
will only discuss spin correlations for the ``completely visible''
decays of the heavy neutral Higgs bosons, into either $\tilde
\chi_1^\pm$ or $\tilde \chi_2^0$ pairs.\s

We can construct three polarization asymmetries from the spins of the
final $\tilde \chi$ states,
\begin{eqnarray}
{\cal A}_a^{ij} = \frac{C^{ij}_a}{C^{ij}_0}\qquad [a=1,2,3].
\end{eqnarray}
We already saw that ${\cal A}_1^{ij}$ is forced to vanish, but the
other CP--odd asymmetry ${\cal A}_3^{ij}$ is allowed. The polarization
asymmetry ${\cal A}^{ij}_2$ is CP--even for neutral Higgs boson
decays, but can yield additional information about the phases. The
statistical error with which an asymmetry can be measured is
proportional to the square root of the number of events in the
sample. The significance with which an asymmetry can be established
experimentally is therefore determined by an effective asymmetry,
defined in terms of the coefficients $C_a^{ij}$ in
eq.(\ref{eq:coefficients}) as
\begin{eqnarray}
\hat{\cal A}_a^{ij} &=& {\cal A}_a^{ij} \sqrt{{\cal B} (H \rightarrow
\tilde \chi_i \tilde \chi_j)}.
\end{eqnarray}
For perfect detection efficiency and polarization analyzing power, the
number of Higgs bosons required for detecting the asymmetry at
1$-\sigma$ level is then simply given by $\hat{A}^{-2}$. The phase
space distribution of $\tilde \chi$ decay products only yields
information about the $\tilde \chi$ spin if the left-- and
right--handed $\tilde \chi$ couplings describing this decay are
different. This is generally true for $\tilde \chi_1^\pm$ decays, so
we expect the analyzing power for $\tilde \chi_1^- \tilde \chi_1^+$
final states to be usually fairly large, in the tens of percent range
at least. On the other hand, the couplings of $\tilde \chi_2^0$ to
both neutral Higgs and neutral gauge bosons give $L$ and $R$ couplings
of equal magnitude. The analyzing power of $\tilde \chi_2^0 \tilde
\chi_2^0$ final states will therefore be very small unless sfermion
exchange contributions are significant. In scenarios with heavy first
and second generation sfermions this might still be the case for
$\tilde \chi_2^0 \rightarrow \tilde \chi_1^0 b \bar b$ and $\tilde
\chi_2^0 \rightarrow \tilde \chi_1^0 \tau^- \tau^+$ decays, which can
also have sizable branching ratios. Obviously, more detailed analyses
would be required to make more precise statements about the analyzing
power, and to estimate the detection efficiencies. Here we simply
present results for the effective asymmetries $\hat{\cal A}$, in order
to show that the asymmetries might in fact be large.\s

\begin{figure}[h]
\vspace*{-2cm}
\centerline{\psfig{figure=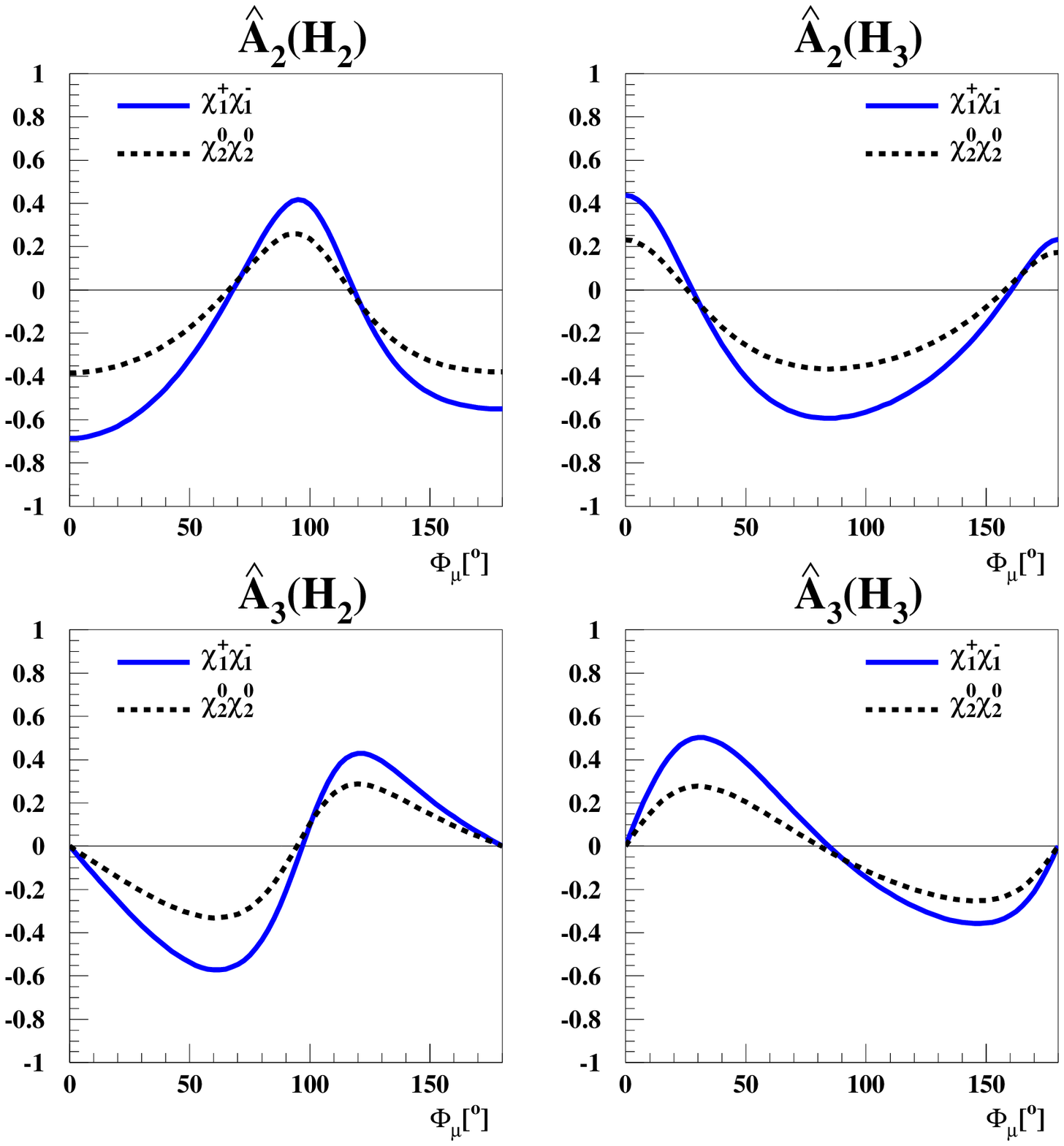,width=15cm}}
\vspace*{-0.9cm}
\caption{\it The polarization asymmetries $\hat{\cal A}_{2,3}$ in the
         supersymmetric decays of the heavy Higgs bosons, $H_2^0$ (left frames)
	 and $H_3^0$ (right frames) with respect to the phase $\Phi_\mu$.
	 The phases $\Phi$ and $\Phi_1$ are set to $0$.}
\label{fig:asym}
\end{figure}

\begin{figure}[h]
\vspace*{-4cm}
\centerline{\psfig{figure=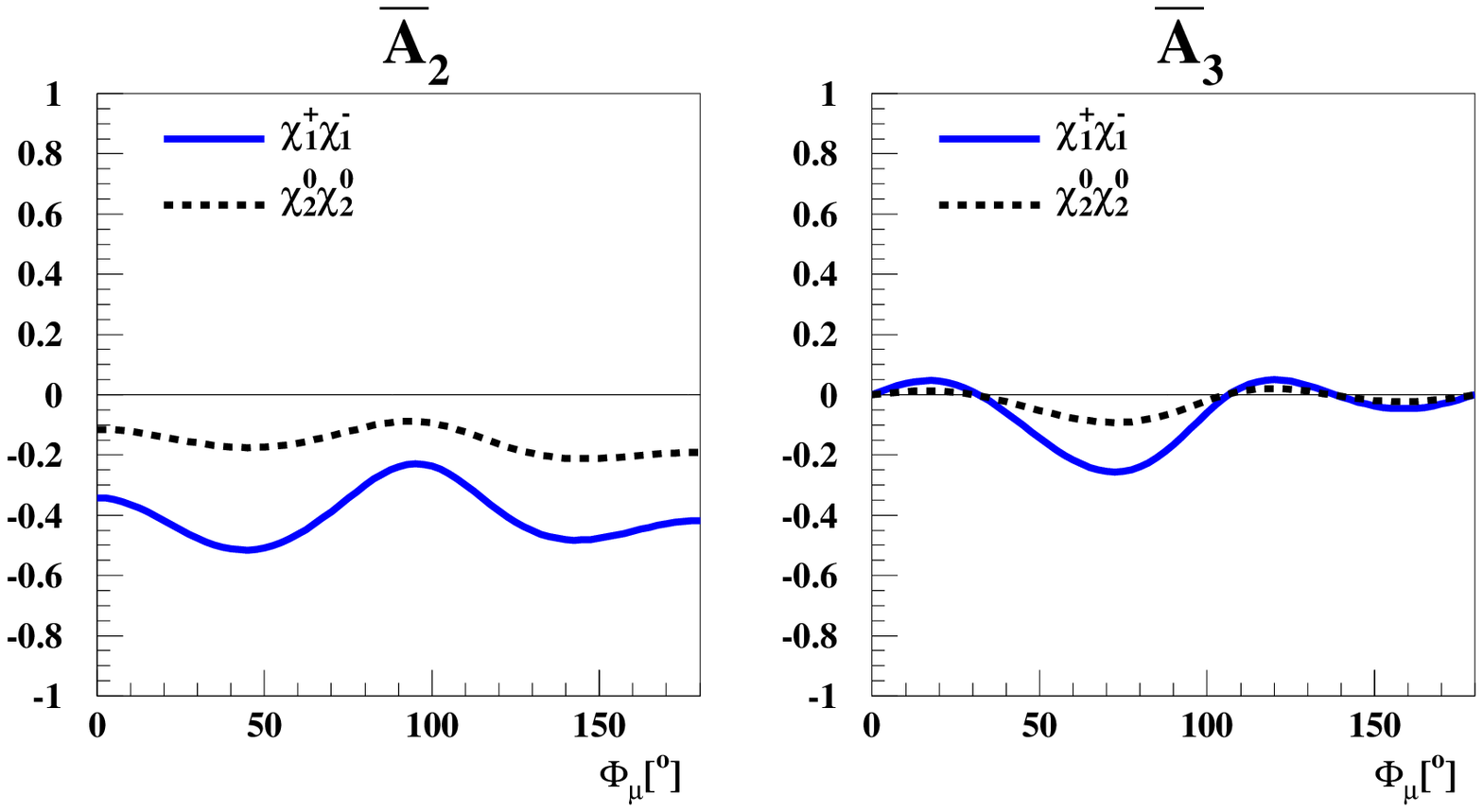,width=15cm}}
\vspace*{-3.9cm}
\caption{\it The average effective polarization asymmetries $\bar{\cal
A}_{2,3}$ in the supersymmetric decays of the heavy Higgs bosons as a
function of the phase $\Phi_\mu$, for $\Phi = \Phi_1 = 0$.}
\label{fig:average}
\end{figure}

In fig.~\ref{fig:asym} we show the effective polarization
asymmetries $\hat{\cal A}_{2,3}$ of the ``completely visible''
supersymmetric heavy Higgs boson decays as functions of $\Phi_\mu$ for
the parameter set (\ref{eq:MCPX}) with $\Phi_1=0$. The left frames
are for $H_2^0$ and the right frames for $H_3^0$. We see that both
asymmetries depend strongly on the phase $\Phi_\mu$ and their sizes
can be significant for a large region of $\Phi_\mu$. Note the strong
anti--correlations between $\hat{\cal A}_i(H_2^0)$ and $\hat{\cal
A}_i(H_3^0)$ for both $i=2$ and $i=3$, which again results from the
different composition of these mass eigenstates in terms of current
eigenstates. Recall that these two heavy neutral Higgs bosons are
almost degenerate. The mass splitting of 2 to 3 GeV in our case should
be sufficient to study $H_2^0$ and $H_3^0$ as separate $s-$channel
resonances at a muon collider \cite{MC1,MC2}. However, it will be
difficult to distinguish decays of $H_2^0$ and $H_3^0$ on an
event--by--event basis at $e^+e^-$ colliders; recall that there the
dominant production process is $e^+ e^- \rightarrow H_2^0 H_3^0$,
i.e., one produces equal numbers of $H_2^0$ and $H_3^0$ bosons. In
fig.~\ref{fig:average} we therefore also show the average effective
asymmetries, defined as
\begin{equation} \label{eq:average}
\bar{\cal A}_i \equiv \frac { {\cal A}_i(H_2^0) B(H_2^0 \rightarrow
\tilde \chi \tilde \chi) + {\cal A}_i(H_3^0) B(H_3^0 \rightarrow
\tilde \chi \tilde \chi) } {\sqrt{ B(H_2^0 \rightarrow \tilde \chi \tilde
\chi) + B(H_3^0 \rightarrow \tilde \chi \tilde \chi)} }.
\end{equation}
We see that averaging in this manner does degrade the asymmetries
significantly; nevertheless, the CP--violating effective asymmetry
might still be of order 20\%.

{\em To summarize.} We studied Higgs boson decays into charginos and
neutralinos in the MSSM with explicit CP violation. The branching
ratios for these decays are sizable whenever the Higgs boson mass
exceeds the sum of gaugino and higgsino masses or whenever there is
significant mixing between gauginos and higgsinos in the light $\tilde
\chi$ states, provided $\tan\beta$ is not very large. We found that
some of these branching ratios depend significantly on the
CP--violating phases.  Much of this sensitivity comes from the
dependence of neutralino\\


\noindent and chargino masses on these phases; these masses can more
easily be measured in the direct production and decay of charginos and
neutralinos. However, the Dirac structure of the relevant coupling
(scalar and/or pseudoscalar) also plays an important role, and is
directly related to CP violation. Moreover, we found that
correlations between the spins of the $\tilde \chi$ states produced in
the decays of heavy neutral Higgs bosons can lead to large
asymmetries, one of which is nonzero only in the presence of
CP violation. This is true even in the absence of CP--violating
mixing between the neutral Higgs bosons, and could thus signal
``direct'' CP violation in Higgs boson decays.  We hope that this
result motivates further detailed investigations, which are needed to
decide whether these large polarization asymmetries are actually
measurable at future colliders.\s

%%%%%%%%%%%%%%%%%%%%%%%%%%%%%%%%%%%%%%%%%%%%%%%%%%%%%%%%
\subsection*{Acknowledgments}
%%%%%%%%%%%%%%%%%%%%%%%%%%%%%%%%%%%%%%%%%%%%%%%%%%%%%%%%

SYC, MD and JS have been partially supported by the Korea Science and
Engineering Foundation (KOSEF) and the Deutsche Forschungsgemeinschaft
(DFG) through the KOSEF--DFG collaboration project, Project No.
20015--111--02--2; SYC was also supported by the Center for High
Energy Physics (CHEP) at Kyungpook National University. The work of
JSL was supported by the Japan Society for the Promotion of Science
(JSPS).\s


\begin{thebibliography}{99}

\bibitem{AP} A. Pilaftsis, Phys. Rev. D {\bf 58}, 096010 (1998);
   Phys. Lett. {\bf B435}, 88 (1998); D.A. Demir, Phys. Rev. D {\bf 60}, 
   055006 (1999); A. Pilaftsis and C.E.M. Wagner, Nucl. Phys. {\bf B553}, 
   3 (1999); M. Carena, J. Ellis, A. Pilaftsis and C.E.M. Wagner, Nucl.
   Phys.  {\bf B586}, 92 (2000).

\bibitem{CDL} S.Y. Choi, M. Drees and J.S. Lee, Phys. Lett. {\bf B481}, 57
   (2000).

\bibitem{EXCP_FC} A. Pilaftsis, Phys. Rev. Lett. {\bf 77}, 4996 (1996);
   K.S. Babu, C. Kolda, J. March--Russell and F. Wilczek, Phys. Rev. 
   {\bf D59}, 016004 (1999); J.F. Gunion and J. Pliszka, Phys. Lett. 
   {\bf B444}, 136 (1998); C.A. Boe, O.M. Ogreid, P. Osland and J. Zhang,
   Eur. Phys. J. {\bf C9}, 413 (1999); B. Grzadkowski, J.F. Gunion and 
   J. Kalinowski, Phys. Rev. {\bf D60}, 075011 (1999); S.Y. Choi and J.S. Lee, 
   Phys. Rev. {\bf D61}, 115002 (2000); {\it ibid}. {\bf 62}, 036005 (2000); 
   S. Bae, Phys. Lett. {\bf B489}, 171 (2000); E. Asakawa, S.Y. Choi, 
   K. Hagiwara and J.S. Lee, Phys. Rev. D 
   {\bf 62}, 155005 (2000); G. Kane and L.-T. Wang, Phys. Lett. {\bf B483}, 
   175 (2000); M. Carena, J. Ellis, A. Pilaftsis and C.E.M. Wagner, Phys.
   Lett. {\bf B495}, 155 (2000); J.S. Lee, hep--ph/0106327; S.Y. Choi,
   K. Hagiwara and J.S. Lee, Phys. Lett. {\bf B529}, 212 (2002).

\bibitem{MC1} S.Y. Choi and J.S. Lee, Phys. Rev. {\bf D61}, 111702  (2000);  
   E. Asakawa, S.Y. Choi and J.S. Lee, Phys. Rev. D {\bf 63}, 115012 
   (2001); M.S. Berger, Phys. Rev. Lett. {\bf 87}, 131801 (2001);
   C. Bl\"{o}chinger et al., hep--ph/0202199.

\bibitem{CL1} S.Y. Choi and J.S. Lee, Phys. Rev. {\bf D61}, 015003 (2000);
   S.Y. Choi, K. Hagiwara and J.S. Lee, {\it ibid.} {\bf 64}, 032004 (2001).

\bibitem{mucol}
S.Y. Choi and M. Drees, Phys. Rev. Lett.  {\bf 81}, 5509 (1998); 
   S.Y. Choi, M. Drees, B. Gaissmaier and J.S. Lee, Phys. Rev. D {\bf 64},
   095009 (2001).

\bibitem{R3a} A. Djouadi, J. Kalinowski and P.M. Zerwas, Z. Phys. C
   {\bf 70}, 437 (1996); {\it ibid.} {\bf 57}, 569 (1993).

\bibitem{DJKZ} A. Djouadi, P. Janot, J. Kalinowski and P.M. Zerwas,
   Phys. Lett. {\bf B376}, 220 (1996).

\bibitem{CKMZ} S.Y. Choi, A. Djouadi, M. Guchait, J. Kalinowski, 
   H.S. Song and P.M. Zerwas, Eur. Phys. J. C {\bf 14}, 535 (2000);
   S.Y. Choi, J. Kalinowski, G. Moortgat--Pick and P.M. Zerwas,
   Eur. Phys. J. C {\bf 22}, 563 (2001).

\bibitem{couploop}
A. Djouadi, M. Drees, P. Fileviez Perez and M. M\"uhlleitner,
hep-ph/0109283 (to appear in Phys. Rev. {\bf D}); H. Eberl, M. Kincel,
W. Majerotto  and Y. Yamada, Nucl. Phys. {\bf B625}, 372 (2002).

\bibitem{IN} T. Ibrahim and P. Nath, Phys. Rev. D {\bf 63}, 035009
(2001), and hep--ph/0204092.

\bibitem{PDG} Particle Data Group, Eur. Phys. J. C {\bf 15} (2000) 1.

\bibitem{Hbb} M. Carena, D. Garcia, U. Nierste and C.E.M. Wagner, Nucl. Phys.
   {\bf B577}, 88 (2000); H. Haber, M. Herrero, H. Logan, S. Penaranda,
   S. Rigolin and D. Themes, Phys. Rev. D {\bf 63}, 055004 (2001);
   H. Eberl, K. Hidaka, S. Kraml. W. Majerotto and Y. Yamada, Phys. Rev.
   D {\bf 62}, 055006 (2000).

\bibitem{french} G. Belanger, F. Boudjema, A. Cottrant, R.M. Godbole and 
   A. Semenov, Phys.Lett. {\bf B519}, 93 (2001).

\bibitem{EDM} For detailed discussions including the constraints from 
   $^{199}$Hg, see T. Falk, K.A. Olive, M. Pospelov and R. Roiban, Nucl. Phys. 
   {\bf B560}, 3 (1999); S. Abel, S. Khalil and O. Lebedev, Nucl. Phys. 
   {\bf B606}, 151 (2000) and references therein.

\bibitem{teslatdr} ECFA/DESY LC Physics Working Group (J.A. Aguilar--Saavedra 
   et al.), TESLA Technical Design Report Part 3, hep--ph/0106315, and 
   references therein. 

\bibitem{virus} O.J.P. \'Eboli and D. Zeppenfeld, Phys. Lett. {\bf B495}, 
   147 (2000).

\bibitem{atlastdr} ATLAS Technical Design Report, Chapter 19, 
   CERN--LHC--99--15, {\tt http://atlasinfo.} \\
   {\tt cern.ch/ATLAS/internal/tdr.html}.

\bibitem{MC2} V.D. Barger, M.S. Berger, J.F. Gunion and T. Han, Phys. Rep.
   {\bf 286}, 1 (1997).

\end{thebibliography}
\end{document}